\documentclass{aa}
\usepackage{times}
\usepackage{epsfig}
\usepackage{colordvi}

\thesaurus{03 %A&A Section 03 : Extragalactic astronomy
	   (11.17.1; %quasars: absorption lines
	   (11.17.4 Q1205-30; %quasars: individual
	    11.06.1)} %Galaxies  : formation

\title{Ly$\alpha$ Emission from a Lyman Limit Absorber at z=3.036
\thanks{Based on observations collected at the European Southern 
Observatory, La Silla, Chile (ESO project No. 60.B-0843).
}
}

\author{J.U. Fynbo \inst{1,2}
   \and B.Thomsen \inst{2}
   \and P. M\o ller \inst{1}}

\institute{European Southern Observatory,  
           Karl-Schwarzschild-Stra\ss e 2,
	   D-85748, Garching by M\"unchen, Germany
           \and
           Institute of Physics and Astronomy,
           \AA rhus University, DK-8000 \AA rhus C.\\
           }

\offprints{J.U. Fynbo}
\mail{jfynbo@obs.aau.dk}

\date{Recieved 17 June 1999/ Accepted }

\begin{document}
\titlerunning{The LLS towards Q1205-30}

\maketitle

\begin{abstract}
Deep, 17.8 hours, narrow band imaging obtained at the ESO 
3.5m New Technology Telescope has revealed extended (galaxy sized) 
Ly$\alpha$ emission from a high redshift Lyman limit absorber. 
The absorber is a z$_{abs} \approx$ z$_{em}$ Lyman limit absorber 
seen in the spectrum of Q1205-30 at z$_{em}$=3.036. The Ly$\alpha$ 
luminosity of the emission line object is 12--14 $\times$ 
10$^{41}$ h$^{-2}$ erg 
s$^{-1}$ for $\Omega_m$=1. The size and morphology of the Ly$\alpha$ 
emitter are both near--identical to those of a previously reported
emission line object associated with a DLA at z=1.934 (Fynbo et al.
1999a), suggesting a close connection between Lyman limit absorbers 
and DLAs.

We also detect six candidate Ly$\alpha$ emitting galaxies in the
surrounding field at projected distances of 156--444h$^{-1}$ kpc
with Ly$\alpha$ luminosities ranging from 3.3 to 9.5 $\times$ 10$^{41}$
h$^{-2}$ erg s$^{-1}$ for $\Omega_m$=1. Assuming no obscuration of
Ly$\alpha$ by dust this corresponds to star formation rates in the range 
0.3 -- 0.9 h$^{-2}$ M$_{\sun}$ yr$^{-1}$. Comparing this to the
known population of high redshift Lyman break galaxies, we find that
the Lyman break galaxies in current ground based samples only 
make up the very bright end of the high redshift galaxy luminosity 
function. A significant, and possibly dominating, population of high 
redshift galaxies are not found in the ground based Lyman break 
surveys.

\end{abstract}

\section{Introduction}

   The amount of information about the galaxy population at high redshift 
($z = 2-4$) has increased tremendously in the last few years. Using the 
Lyman-break technique several hundred high redshift star forming galaxies, 
Lyman-break galaxies (LBGs), have been detected and studied with imaging 
and spectroscopy (Steidel et al. 1996). It is, however, not yet known how 
complete the Lyman-break technique is in detecting high redshift galaxies.

   An independent route along which to study the galaxy population at high 
redshift is via the high column density QSO absorption lines systems. 
The advantage of high column density QSO absorption lines systems is that 
a wealth of information on the
chemical evolution can be and has been obtained by studying the metallicity
and dust content of the absorbers from high resolution spectroscopic
studies of the background QSOs (e.g. Lu et al. 1996). However, the
spectroscopic studies will not tell us anything about the emission
properties of the absorbing galaxies.

  Only when combining the information obtained from the
LBG-studies (e.g. the luminosity function of LBGs), the absorption line
statistics for QSO absorption lines systems and the properties of galaxy 
counterparts of QSO absorption lines systems
can we hope to disentangle the observational selection biases which
each of the different studies suffer from and obtain a more complete insight
into the nature of the high redshift galaxy population.

The absorption line systems with the highest HI column density,
$N(HI)>2\times10^{20}$ cm$^{-2}$, are the Damped Ly$\alpha$ Absorbers 
(DLAs, Wolfe et al. 1986)\footnote{When comparing absorbers to
statistical samples of DLA absorbers, it is important that they
meet this N(HI) criterion. However, some authors also refer to 
absorbers with $N(HI)<2\times10^{20}$ cm$^{-2}$ as DLAs, e.g. Lanzetta et 
al. 1997.}. Such high HI column densities are at low redshift only found  
in the disks of spiral galaxies. It it also interesting to note, that
active star formation in spiral galaxies only occurs when the HI column
density of the disk exceeds $2\times10^{20} cm^{-2}$ (Kennicutt 1989).
For these reasons, DLAs are widely believed to trace the central parts 
of forming galaxies. Much telescope time has been dedicated to the 
narrow band imaging of DLAs over the past decade 
(e.g. Lowenthal et al. 1995 and references therein), but so far
resulting in only two confirmed detections for the DLAs towards
PKS0528-250 (M\o ller \& Warren 1993a,b; Warren \& M\o ller 1996),
and Q0151+048A (M\o ller et al. 1998a; Fynbo et al. 1999a)
In addition a spectroscopically confirmed broad band detection
of the DLA towards DMS2247-0209 (Djorgovski 1998) and a purely 
spectroscopic detection of Ly$\alpha$ emission from the DLA towards
Q2059-360 have been reported (Pettini et al. 1995, 
Leibundgut \& Robertson 1998), as well as a number of DLA
candidates, which have not yet been confirmed by spectroscopy 
(Steidel et al. 1994, 1995, Arag\'on-Salamanca et al. 1996, 
LeBrun et al. 1997).

   QSO absorption line systems with N(HI) larger than a few times 
$10^{17}$ cm$^{-2}$ are optically thick at the Lyman limit and refered to
as Lyman Limit Systems (LLSs). Lyman limit absorption is believed to 
occur in extended gaseous haloes of galaxies, because the neutral
hydrogen column density is much larger than in the intergalactic
medium and the gas is not predominantly neutral as in DLAs. At low
redshifts, where the 
Lyman limit cannot be observed from the ground, LLS are thought to be 
traced by MgII absorption, because MgII absorption only occurs in
optically thick clouds (Schaeffer 1983). The extensive study of the galaxy
counterparts of MgII absorbers presented in Guillemin \& Bergeron
(1997) shows that MgII absorption predominantly occurs in  
Sbc or Scd galaxies, but that the objects range from ellipticals to
irregular galaxies. At high redshifts the Lyman Break technique were 
originally used to look for galaxies responsible for Lyman limit 
absorption in QSO spectra (Steidel \& Hamilton 1992). Two candidates
were found in six fields (Steidel et al. 1995), but for
only one of these, the LLS towards Q2233+131, has confirming 
spectroscopy been published (Djorgovski et al. 1996).

Three out of the four confirmed high redshift DLAs that have been 
detected in emission at present are at approximately the same redshift 
as the background QSO. In order to examine whether
$z_{abs} \approx z_{em}$ systems indeed are more active emitters
(e.g. due to induced star formation or to photoionisation by the QSO;
for the suggested mechanisms see M{\o}ller \& Warren 1993b and
Fynbo et al. 1998a)
we have initiated a programme aimed at studying the galaxy counterparts
of this special subgroup of high column density QSO absorption line
systems. As part of this programme we chose to study the quasar
Q1205-30 of which a spectrum published by Lanzetta et al. (1991)
shows the presence of a strong LLS close to the emission redshift of
the QSO. The redshift of the quasar is $z_{em} = 3.036$.

In Sect. 2 of this paper we describe the observations obtained with
the ESO New Technology Telescope (NTT) and the basic data reduction. In
Sect. 3 we discuss the photometry, the selection of Ly$\alpha$
emission line candidates, and the correction for the quasar point
spread function. In Sect. 4 we discuss our results.
We adopt a Hubble constant of 100h$^{-1}$ km s$^{-1}$ Mpc$^{-1}$ and
assume $\Omega_m$=1 and $\Omega_{\Lambda}$=0 unless otherwise stated.

\section{Observations and Data Reduction}
A spectrum of the quasar Q1205-30 can be found in Lanzetta et al. 
(1991), but its celestial coordinates were never made publicly 
available. To obtain an image we therefore had to ``rediscover'' 
Q1205-30 on a copy of the prism plate UJ9085P kindly made available 
to us from the UK Schmidt Telescope Objective Prism Survey (for 
details of the procedure see Fynbo et al. 1999b). 
Celestial coordinates of Q1205-30 are RA 12 05 35.72, 
Dec -30 14 25.8 (1950).

Deep imaging of the field was subsequently carried out in NTT service
mode (Silva \& Quinn 1997, Silva 1998, Woudt \& Silva 1999). The 
field was imaged through
standard B and Bessel I filters, as well as through a special
20\AA \ (fwhm) narrow band filter manufactured by Custom Scientific.
The narrow band filter (CS 4906/20) is centred at 4906\AA, which
is the wavelength of redshifted Ly$\alpha$ at $z=3.036$. The service
mode data were obtained with the ESO Multi-Mode Instrument (EMMI) on
the NTT during several nights of January, February and 
March, 1998. The CCD used in the red arm of EMMI was a SITe TK2048
with a pixel scale of 0.27 arcsec. The blue arm CCD was a SITe TK1024
with a pixel scale of 0.35 arcsec. The total integration
times in each filter and the seeing in each of the combined frames are 
given in Table 1.

%=====================Begin Table 1=================================
\begin{table}
\begin{center}
\begin{tabular}{cccc}
\hline\noalign{\smallskip}
\multicolumn{1}{c}{Filter}& Combined fwhm & Exposure Time & Pixel scale \\
           & (arcsec) & (sec) & (arcsec/pixel)    \\
\noalign{\smallskip}
\hline\noalign{\smallskip}
CS 4906/20 & 1.4  & 64000 & 0.35 \\
B          & 1.3  &  3600 & 0.35 \\
I          & 0.96 &  5600 & 0.27 \\
\noalign{\smallskip}
\hline
\end{tabular}
\caption{Journal of observations, NTT, January through March 1998.}
\end{center}
\label{log}
\end{table}
%=====================End Table 1=====================================

   Landolt (1992) photometric standards and three spectrophotometric
standard stars Eggr99, Feige 56 and L970 were used for the photometric
calibrations.

I--band imaging was obtained with the red arm of EMMI. The CCD in
the red arm of EMMI can be read in either single port (D only) or
dual port (A + D) read--out mode.
Unfortunately it showed up that for our programme a mix of single and
dual port read--out had been employed, and we had to follow a somewhat
complex reduction procedure to make up for this.

The disadvantage of the single port read--out is the longer read time
required (which is particularly a problem while obtaining twilight
flats). The disadvantage of the dual port read--out is that the bias
level, the gain (and drift of the gain), and the readout
noise are different for the two image sections. Also the locations of
traps (bad columns) depend on the direction of charge transfer along the
columns. In particular the section read through port A has more bad
columns than the same section read through port D.
The I--band science exposures of Q1205-30 were read out using dual port
read--out, whereas the exposures of standard stars were read out using
single port mode
only. This mixed mode operation requires 2 sets of biases,
as well as two sets of flat fields, and two different ways of reducing
the images. Twilight flats were only obtained for single port mode
(appropriate for the standard star exposures), while dome flat were
taken with dual port read--out (appropriate for the science frames).
The dual read exposures were treated as if the
A and D subsections were two individual frames. Finally, bias frames,
flat field exposures, and science exposures were pasted together to form
2048x2046 frames. The subsequent flat-fielding was done in the usual
way.

The CCD in the blue arm of EMMI can be read in single port mode only.
All science frames obtained with the blue arm of EMMI (all B--band and
narrow band data) were
bias--subtracted and flat--fielded using standard routines.

Following the basic reduction the images in each of the three filters
were combined employing the code described in M\o ller \& Warren
(1993a). This code optimizes the signal--to--noise for faint objects
in the field, for which the noise is well understood via propagation of
read--out--noise and photon shot noise.

All magnitudes quoted in this paper are on the AB system. The
narrow band data were calibrated directly onto the AB system using the
spectrophotometric standard stars, and these magnitudes are denoted 
$n(AB)$. We found that the
colour terms for both the I and B filters were consistent with zero, 
hence we use the equations $I(AB)=I+0.43$ and 
$B(AB)=B-0.14$ (Fukugita et al. 1995) to put the broad band magnitudes 
onto the AB system. Details of the sky brightness and sky noise in the
combined images are provided in Table 2.

%=====================Begin Table 2====================================
\begin{table}
 \begin{center}
 \caption{Measured level and rms of sky surface brightness.}
 \begin{tabular}{@{}lcccccc}
  passband & sky                & rms SB\\
  \hline
           & mag$_{AB}$. arcsec$^{-2}$ & mag$_{AB}$. arcsec$^{-2}$ \\
  \hline
  n(AB)    & 21.9--22.4 & 27.9 \\
  B(AB)    & 22.5       & 28.5 \\
  I(AB)    & 19.7       & 27.5 \\
  \hline
 \end{tabular}
 \label{sky}
 \end{center}
\end{table}
%=====================End Table 2====================================

We reach limiting magnitudes (5$\sigma$) of n(AB)=25.0, B(AB)=25.8 and
I(AB)=25.1. A narrow band AB magnitude of 25.0 corresponds 
to a Ly$\alpha$ flux of $1.1\times10^{-17}$ erg \ s$^{-1}$ cm$^{-2}$.

\section{Results}
\begin{figure}[t]
 \epsfig{file=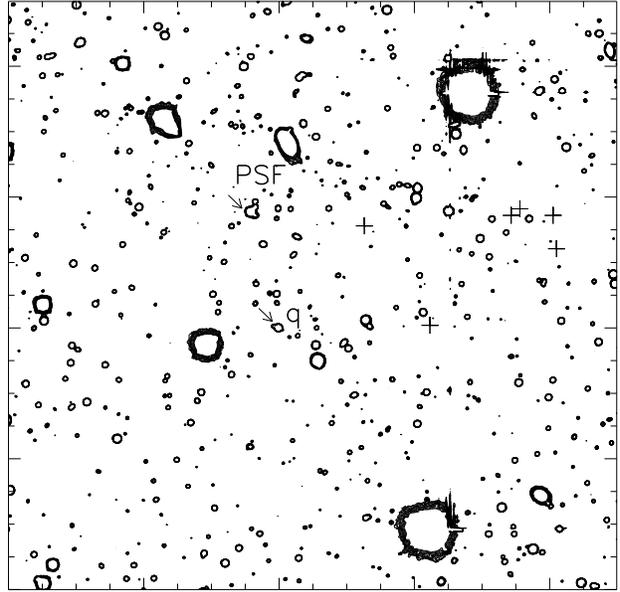,width=8.5cm} 
 \caption{Contour plot of the I--band frame showing
 the 315x315 arcsec$^{2}$ field surrounding Q1205-30. 
 North is up and east to the left. Q1205-30 is shown by the arrow next
 to the ``q''. The star used
 to define the PSF is marked `PSF' (see Sect.~\ref{psfsub}). 
 The positions of six candidate Ly$\alpha$ emitting galaxies are 
 marked with a ``+'' (see Sect.~\ref{field2}).
 }
\label{field}
\end{figure}

A contour plot of the combined I--band image of the 315x315 arcsec$^2$
field surrounding Q1205-30 is shown in Fig.~\ref{field}.
Q1205-30 itself is here marked by a `q' and an arrow. The star
from which the central core of the Point Spread Function was defined
(see Sect.~\ref{psfsub}) is marked `PSF'. Six candidate
Ly$\alpha$ emitting galaxies (see Sect.~\ref{field2}) are marked
``+''.

%=====================Begin Figure 1====================================
%=====================End Figure 1====================================

\subsection{Objects near the line of sight towards Q1205-30}
\label{psfsub}

Q1205-30 was selected for observation because of the high column
density absorber seen along the line--of--sight towards it. Previous
detections of high column density absorbers (for a recent summary see
M{\o}ller \& Warren 1998) suggest that their absorption cross--section
is very small, and that emission from the object therefore likely will
be hidden under the quasar PSF. To search for emission from the
absorbing object we therefore performed a careful PSF subtraction as
detailed in this section.

The basic principles of the PSF subtraction are as described in
M{\o}ller \& Warren (1993a), and Fynbo et al. (1999a).
For a first approximation to the PSF we used the star marked `PSF' in
Fig.~\ref{field}. This is the nearest unsaturated point source
significantly brighter than Q1205-30, but as is evident from
Fig.~\ref{field} it has four nearby neighbour objects (projected
distances in the range 4--9 arcsec). The signature of those four
objects does not affect the core of the PSF. Their effect on the
halo of the PSF was masked out and removed via substitution of the
masked regions by areas selected from scaled high S/N halos from
isolated point--sources brighter than the PSF--star (and hence saturated
in the central core). The comparatively
large projected distances (4 arcsec and larger) means that the presence
of those objects, even if not corrected for, would not in any case
have had any effect for the results reported below.

The DAOPHOT-II (Stetson 1997) extension ALLSTAR was used to perform
the final PSF model fit and subtraction. Identical procedures
were followed for the PSF subtraction in all combined images:
Narrow band, B and I.

%=====================Begin Figure 2====================================
\begin{figure}
 \epsfig{file=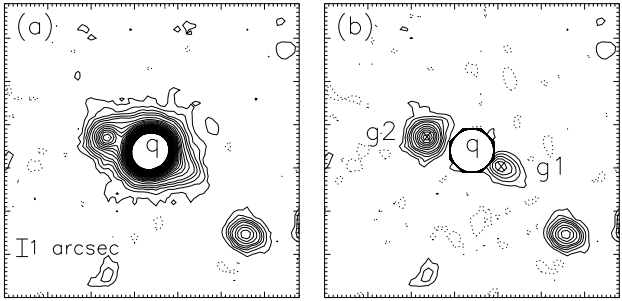,width=8.5cm} \caption{Contour plot of the
 I--band frame showing an 18x18 arcsec$^{2}$ field
 surrounding Q1205-30. North is up and east to the left. The
 frame has been smoothed by a 3x3 pixel boxcar filter.
 {\it Left:\,} final combined frame (QSO is marked q).
 The contour levels are at -9, -6, -3, 3, 6, 9... in units of the
 background noise, with the dotted contours being negative.
 {\it Right:\,} same as left but here after subtraction of the quasar
 PSF, revealing two galaxies marked g1 and g2 at separations 2.2
 and 2.8 arcsec (from the QSO) respectively. Their centroids are
 marked ``$\times$''.
 A circle of radius 1.3 arcsec has been drawn
 around the centre of the subtracted PSF (see text for details).}
\label{figI}
\end{figure}
%=====================End Figure 2====================================

%=====================Begin Figure 3====================================
\begin{figure}[ht]
 \epsfig{file=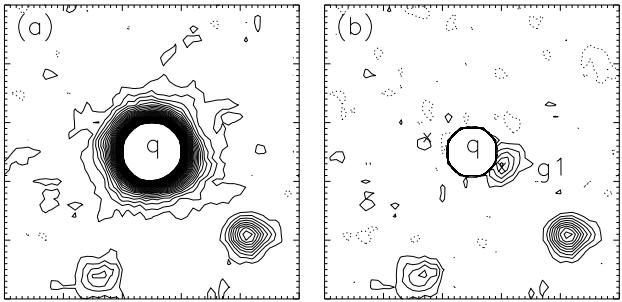,width=8.5cm} \caption{Same as Fig.2 but here
 for the B--band image.
 The contour levels are at -6, -4, -2,  2, 4, 6... in units of the
 background noise with dotted contours being negative.
 {\it Right:\,} Here a circle of radius 1.5 arcsec has been drawn
 around the centre of the subtracted PSF. The ``$\times$''s mark the
 centroids of the galaxies g1 and g2 as measured on the I--band image.
 }
\label{figB}
\end{figure}
%=====================End Figure 3====================================

In Fig.~\ref{figI} we show a 18x18 arcsec$^2$ cut--out of the I--band
image centred on Q1205-30 before (a) and after (b) subtraction of the
quasar PSF. The quasar is marked ``q'', and it is clearly seen that two
faint galaxies (named g1 and g2) were blended with the quasar PSF.
In Fig.~\ref{figI}(b) we have drawn a circle of radius 1.3 arcsec
around the centre of the subtracted PSF. Inside this circle the large
residuals (not shown) from the PSF subtraction make it impossible to
search for objects. The galaxy g2 is well separated from the
PSF--subtraction residuals so its projected distance from Q1205-30
(2.8 arcsec) is well determined. In contrast, g1 is partly embedded in
the residuals. The galaxy g1 could therefore in reality be an elongated
object extending across the quasar, and the {\it measured} projected
distance (2.2 arcsec) is hence an upper limit to the {\it true}
projected distance. The measured I--band centroids of the galaxies
g1 and g2 are marked by an ``$\times$'' in Fig.~\ref{figI}(b).

In Fig.~\ref{figB}(a,b) we show the same 18x18 arcsec$^2$ cut--out
of the B--band image. Here we have drawn a circle of radius 1.5 arcsec,
again to mask out the area where PSF subtraction makes it impossible to
search for objects. The ``$\times$''s here again mark the centroids of the
galaxies g1 and g2 as measured on the I--band image. In the B--band
image g1 is found at the same projected distance as in the I--band
image while the extremely red object g2 is not detected.
 
%=====================Begin Figure 4====================================
\begin{figure}[ht]
 \epsfig{file=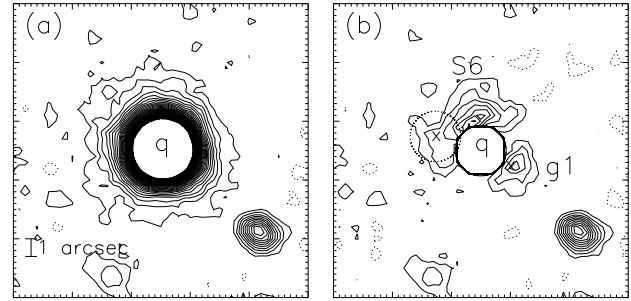,width=8.5cm} \caption{
{\it Left:\,} Same as Fig.2 but here for the narrow band image. The 
contour levels are at -6, -4, -2, 2, 4, 6..., in units of the 
background noise, with the dotted contours being negative. 
{\it Right:\,} Here the PSF of the quasar has been subtracted,
revealing a large extended object (named S6) to the north, possibly
extending to the east, of Q1205-30. A circle of radius 1.5 arcsec has 
been drawn around the centre of the subtracted PSF. The dotted circle
marks the extend of g2 in the I--band image, and the ``$\times$''s 
mark the centroids of the galaxies g1 and g2 as measured on the I--band 
image.
}
\label{fignarrow}
\end{figure}
%=====================End Figure 4====================================

%=====================Begin Figure 5====================================
\begin{figure}[ht]
 \epsfig{file=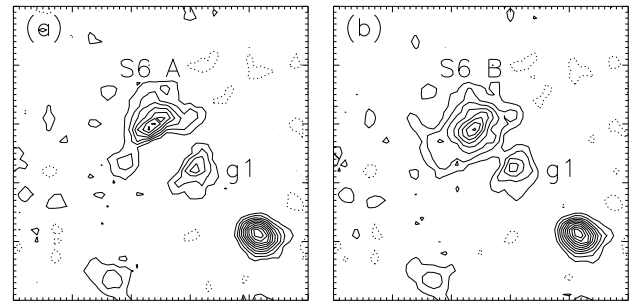,width=8.5cm} \caption{Two ``extreme'' models
 of S6. {\it Left:\,} in model A we assume that all the flux seen at
 the position of g2 is indeed from g2, and we make no allowance for
 flux from S6 within the central core of the quasar PSF.
 {\it Right:\,} in model B we assume that no significant flux is seen
 from g2, and we model S6 within the central core of the quasar PSF
 assuming that S6 is symmetric about its major axis with position
 angle 120$^\circ$ east of north.
 }
\label{fignarrow2}
\end{figure}
%=====================End Figure 5====================================

In Fig.~\ref{fignarrow}(a,b) we again show the same 18x18 arcsec$^2$
cut--out as above, but here from the combined narrow band frame.
The ``$\times$''s here again mark the centroids of the
galaxies g1 and g2 as measured on the I--band image. In this image we
clearly see an extended object (named S6) to the north of the quasar,
possibly extending all the way to the east of the quasar. Note that
g1 here is found at a slightly larger projected distance from the
quasar than in the I--band and B--band images. This illustrates the
point we made above that the PSF subtraction will tend to ``push''
the object out at a larger distance, because part of the object is
embedded in the non--recoverable central part of the PSF. This will
then also impact the photometry, as the total flux of the object will
be underestimated. The position of g2 (marked by '$\times$' and a dotted
circle) corresponds to a faint protrusion on the extended object. 
Since the narrow band is within the B passband, we would indeed 
expect to see only very weak emission from the extremely red object 
g2.

%=====================Begin Table 3=================================
\begin{table}
\begin{center}
\caption{Photometric properties of the QSO, of the faint
galaxies g1 and g2 and of the Ly$\alpha$ emitter S6. Upper 
limits to magnitudes are 2$\sigma$ except for g1, where the
limit is determined from model A mentioned below.}
\begin{tabular}{l r r rrrr}
Object & B(AB)           & I(AB)      & n(AB)      & Apert.\\
       &                 &            &            & (arcsec$^2$)\\
\hline
q    & 18.92        & 18.13          & 18.28             & 60 \\
S6 A & $>26.2$      & $>25.2$        & 22.90$\pm0.10$    & 19 \\
S6 B & $>26.2$      & $>25.2$        & 23.10$\pm0.10$    & 19 \\
g1   & $24.8\pm0.1$ & $23.3\pm0.1$   & $24.1\pm0.2$      & 9.6 \\
g2   & $>26.5$      & $22.22\pm0.05$ & $\geq24.8\pm0.3$  & 9.6 \\
\hline
\end{tabular}
\end{center}
\label{photometry}
\end{table}
%=====================End Table 3=====================================

\subsection{Photometry of objects near the QSO line of sight.}
\label{photnear}

   In this subsection we describe the photometry on Q1205-30, g1, g2 and
S6 in the three filters B, I and narrow band. For g1 we measured the
flux inside a circular aperture of diameter 3.5 arcsec. The resulting
magnitudes presented in Table 3 will be somewhat underestimated since
we miss the flux closer than 1.5 arcsec from the QSO.
For g2 we also measured fluxes in circular apertures of diameter 
3.5 arcsec. We did not detect g2 in B, and hence provide the 2$\sigma$
detection limit. The galaxy g2 is very red, consistent with being an
old stellar population at a redshift of about 0.5 or more. In
Sect.~\ref{discussLLS} we discuss the possible effects of g2 on the
line of sight due to gravitational lensing. In the narrow band image 
we can not in an objective way determine whether the
flux detected at the position of g2 originates from g2 or from S6.
Hence we chose to consider two extreme cases for the photometry of
S6 and g2 in the narrow filter.

In model A (minimum Ly$\alpha$ flux model; Fig~\ref{fignarrow}a)
we subtracted the maximum flux we can possibly
ascribe to g2 in the narrow band. We used a model of g2 obtained by
smoothing the I band image of g2 to the seeing of the narrow band image,
and then determined the maximal scaling of this model which
after subtraction left residuals consistent with the noise. The
remaining flux, making no correction for flux from S6 within the
central core of the subtracted quasar PSF, was assigned to S6.
In model B (maximum Ly$\alpha$ flux model; Fig~\ref{fignarrow}b)
we assumed that all the flux seen north and east of
Q1205-30 originates from S6 and not from g2. In order to estimate
the flux within the central core of the quasar PSF we assumed that
S6 is symmetric about its major axis (PA 120$^\circ$ east of north).
A model of S6 was made as follows. First we flipped the image of S6
around its major axis. In the PSF subtracted image we then replaced
the region inside radius 1.5 arcsec from the quasar PSF with the
flipped image. The flux of the model was measured using a circular
aperture of diameter 4.9 arcsec.

The impact parameter of S6 was found to be 1.8 arcsec in model A and
1.5 arcsec in model B.
The Ly$\alpha$ flux of S6 is 6.6$\pm$0.6$\times10^{-17}$ ergs
s$^{-1}$ cm$^{-2}$ (model A) and 7.9$\pm$0.7$\times10^{-17}$ ergs
s$^{-1}$ cm$^{-2}$ (model B).

\subsection{Candidate Ly$\alpha$ emitting galaxies in the field}
\label{field2}

%=====================Begin Figure 6====================================
\begin{figure}[t]
 \begin{center} \epsfig{file=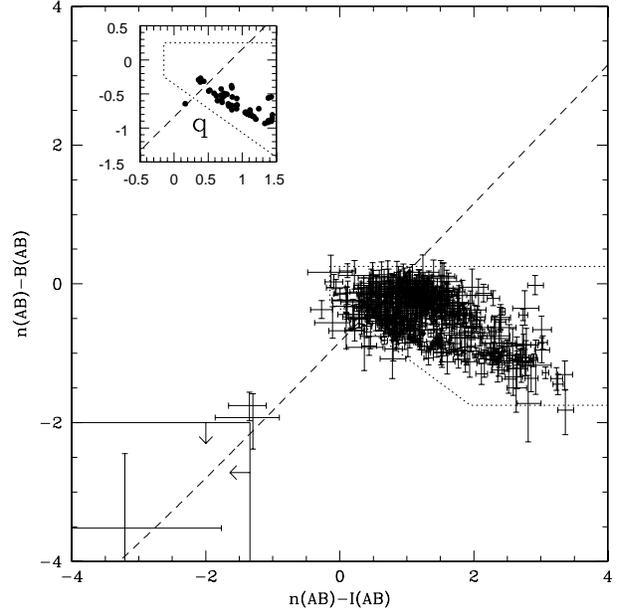,width=8.5cm} \end{center}
 \caption{
 Two-colour diagram $n(AB)-B(AB)$ {\it versus} $n(AB)-I(AB)$
 for objects detected at a S/N $>5$ in the
 narrow band frame. The dashed line is a line of constant
 $B-I$ in the continuum (chosen to be that of Q1205-30), but with a
 spectral feature in the narrow filter ranging from a strong absorption
 line in the upper right hand corner to a strong emission line in the
 lower left hand corner. Three of the six candidate emission line
 galaxies are seen in the lower left corner. The remaining three are
 detected in neither B nor I. For these galaxies the lines with arrows 
 indicate the 2$\sigma$ allowed region in the diagram. The dotted 
 lines confine the expected region of objects with no special features 
 in the narrow filter. The inserted diagram shows the region of the 
 plot containing Q1205-30 (marked q) and objects with S/N$>30$.
}
\label{fig4}
\end{figure}
%=====================End Figure 6====================================

The ``narrow minus on--band--broad'' versus ``narrow minus
off--band--broad'' colour/colour plot technique (M{\o}ller \& Warren,
1993a; Fynbo et al. 1999a) has proven a powerful tool to identify
galaxies in the faintest end of the high redshift galaxy luminosity
function (LF). We here describe the search procedure and the photometric
measurements carried out with the aim to produce the plot shown in
Fig.~\ref{fig4}.

Identification and photometric measurements of objects in the field
was done with the photometry package SExtractor (Bertin and Arnouts,
1996). We used a minimum object extraction area of 8 
pixels and a detection threshold of 1.3$\sigma$ above the background.
As our detection image we selected to use a weighted average of the
three combined frames. The weights were chosen to be the inverse of the
variance in each of the combined frames. As a detection filter we used 
a Gaussian filter with a full-width-at-half-maximum (fwhm) of 4 pixels 
similar to the fwhm of point sources in the detection image, which is
3.4 pixels.

In total we detected 473 objects with a signal--to--noise ratio (S/N) 
$>5$ in the narrow band. In Fig.~\ref{fig4} we show the colour--colour 
diagram $n(AB)-B(AB)$ {\it versus} $n(AB)-I(AB)$ for the entire sample. 
In this diagram objects with Ly$\alpha$ emission in the narrow band
filter will be located in the lower left corner while objects with
absorption in the narrow filter will be located in the upper right hand
corner. Since our narrow filter is centred at 4906\AA, which is in the
red wing of the B filter, the $n(AB)-B(AB)$ colour of an object will 
depend on the slope of the objects spectral energy distribution. Very
red objects will, therefore,
appear to have a slight excess emission in the narrow band, causing
the locus of continuum sources to be slightly tilted towards the lower
right of the diagram.

We determined the expected region for objects with no special feature in
the narrow band by calculating the position in this diagram for a wide
range of galaxy spectral energy distributions taken from the models
of Bruzual \& Charlot (1993) 
The galaxy models were
calculated for several redshifts in the range $z=0$ to $z=3.5$, and
were corrected for Ly$\alpha$ line blanketing due to intervening
absorbers (M{\o}ller \& Jakobsen 1990).
The resulting region is indicated by the dotted line in Fig.~\ref{fig4}.

Inspection of Fig.~\ref{fig4} clearly shows that the vast majority of
the detected objects indeed conform to the predicted colours. Three
objects to the lower left are, however, found to lie significantly
outside the locus of continuum objects. All three are found
in the region expected for blue objects with an emission line 
in the narrow filter.

The detection limit in the combined image used for object detection
is dominated by the deeper broad band images. Faint objects with large
equivalent width Ly$\alpha$ emission might therefore be missed by the
detection algorithm. To make up for this we repeated the detection
procedure, but this time using the narrow band image for detection.
SExtractor found six emission line objects with S/N$>5$. Three of those
were the ones reported already from the combined image detection,
while the other three remain undetected in broad band.

\begin{figure}[ht]
\begin{center}
 \epsfig{file=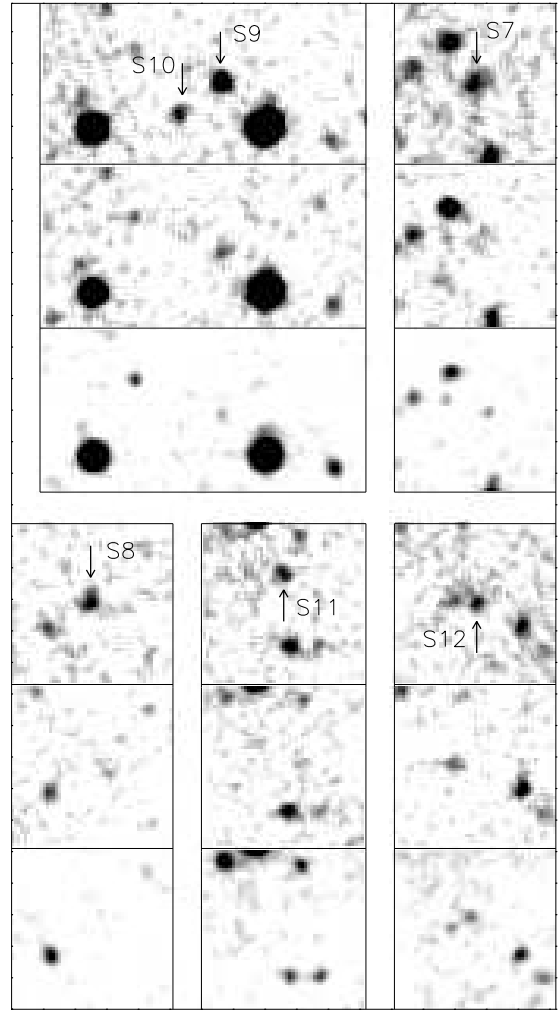,width=7.21cm} 
 \caption{Extractions from the combined narrow band (top), B band (middle) 
 and I band (bottom) for each of the six candidate emission line galaxies 
 detected with S/N$>5$ (see Sect.~\ref{field2}). In each field emission 
 line galaxies have been marked with an arrow. The size of the square
 fields is 18x18 arcsec$^{2}$.
 }
\label{extracts}
\end{center}
\end{figure}

In addition to S6 we hence detect a total of six emission line objects
at S/N$>5$. The positions of the six objects are marked by crosses in
Fig.~\ref{field}. Image cut--outs showing the 6 objects in each of
the three bands are reproduced in Fig.~\ref{extracts}.
Photometric properties of the emission line objects are reported in
Table 4. Magnitudes were measured using both SExtractor isophote
apertures and large circular apertures. Emission line fluxes 
corresponding to the measured n(AB) aperture magnitudes range from 
$1.9\times10^{-17}$ erg s$^{-1}$ cm$^{-2}$ to $5.4\times10^{-17}$ erg 
s$^{-1}$ cm$^{-2}$.

%=====================Begin Figure 7====================================
%=====================End Figure 7====================================

%=====================Begin Table 4=================================
\begin{table*}
\begin{center}
\caption{Photometric properties of the six candidate Ly$\alpha$
emitting galaxies. Upper limits to magnitudes and lower limits to
equivalent widths (EWs) are 2$\sigma$. SFRs are calculated assuming a 
$\Omega_m$=1 and $\Omega_{\Lambda}$=0 universe, and using the 
prescription SFR=L(Ly$\alpha$)/1.12 $\times$ 10$^{42}$ erg s$^{-1}$.}
\begin{tabular}{l rrrccc}
Object & n(AB)           & B(AB)      & I(AB)      & Ly$\alpha$
 flux  & EW$_{rest}$ & SFR  \\
\hline
Isophote &               &            & & 10$^{-17}$ ergs 
s$^{-1}$ cm$^{-2}$ & \AA \ & h$^{-2}$ M$_{\sun}$ \\
\hline
S7       & 24.13$\pm$0.14 & 25.8$\pm$0.3 & 25.0$\pm$0.3 
& 2.56$\pm$0.33 & 20$\pm7$ & 0.40$\pm$0.05 \\
S8       & 24.46$\pm$0.15 & $>$26.9      & $>$26.0      
& 1.89$\pm$0.26 & $>$45  & 0.30$\pm$0.04 \\
S9       & 23.43$\pm$0.09 & 25.3$\pm$0.2 & 24.9$\pm$0.3 
& 4.88$\pm$0.40 & 25$\pm$6 & 0.77$\pm$0.06 \\
S10      & 24.78$\pm$0.18 & $>$27.0      & $>$26.1      
& 1.41$\pm$0.23 & $>$36  & 0.22$\pm$0.04 \\
S11      & 24.83$\pm$0.18 & $>$27.0      & $>$26.2      
& 1.34$\pm$0.22 & $>$34  & 0.21$\pm$0.04 \\
S12      & 24.96$\pm$0.19 & $>$27.1      & $>$26.3      
& 1.19$\pm$0.21 & $>$33  & 0.19$\pm$0.03 \\
\hline
Aperture & (3.5 $\arcsec$ diameter) &            & \\
\hline
S7       & 23.90$\pm0.13$ & 25.5$\pm$0.3 & 25.0$\pm$0.3 
& 3.17$\pm$0.39 & 18$\pm$6 & 0.50$\pm$0.06 \\
S8       & 23.90$\pm0.13$ & $>$26.1      & $>$25.6      
& 3.16$\pm$0.39 & $>$35 & 0.50$\pm$0.06 \\
S9       & 23.32$\pm0.07$ & 25.2$\pm$0.2 & 24.6$\pm$0.2 
& 5.40$\pm$0.37 & 25$\pm$6 & 0.85$\pm$0.06 \\
S10      & 24.33$\pm0.19$ & $>$26.1      & $>$25.6      
& 2.13$\pm$0.42 & $>$22 & 0.34$\pm$0.07 \\
S11      & 24.42$\pm0.21$ & $>$26.1      & $>$25.6      
& 1.96$\pm$0.42 & $>$20 & 0.31$\pm$0.07 \\
S12      & 24.45$\pm0.22$ & $>$26.1      & $>$25.6      
& 1.90$\pm$0.42 & $>$19 & 0.30$\pm$0.07 \\
\hline
\end{tabular}
\end{center}
\label{candphot}
\end{table*}
%=====================End Table 4=====================================

   As seen in the insert in the upper left corner of Fig.~\ref{fig4},
Q1205-30 has only a slight excess emission in the narrow band filter
which covers Ly$\alpha$ of the quasar also. The weakness of this
excess emission is the result of the blue wing of the Ly$\alpha$
emission line of the QSO being absorbed partly by the Lyman 
limit absorber.

\section{Discussion and Conclusions}
\subsection{The Lyman Limit Absorber}
\label{discussLLS}

We have detected an object, S6, with extended line emission north 
and east of the quasar Q1205-30. The emission line of S6 is detected 
in a narrow band tuned to Ly$\alpha$ of a Lyman Limit absorber in 
front of Q1205-30, and the most likely interpretation is, that we see 
the Lyman Limit absorber in Ly$\alpha$ emission. Spectroscopic 
follow--up is required to confirm this 
tentative conclusion. The large residuals from the PSF subtraction
of the quasar introduce errors in both the total line luminosity
observed, and the impact parameter. We have therefore considered two 
extreme cases described in Sect.~\ref{photnear}; the minimal 
Ly$\alpha$ flux case (model A) and the maximal Ly$\alpha$ flux case
(model B). For the Ly$\alpha$ 
luminosity and the impact parameter in the two models we find 
L=11.6$\pm$1.1 $\times$ 10$^{41}$ h$^{-2}$ erg s$^{-1}$(14.1$\pm$1.3 
$\times$ 
10$^{41}$ h$^{-2}$ erg s$^{-1}$) and d=6.5 h$^{-1}$ kpc (5.4 h$^{-1}$ 
kpc) for model A (model B). In the following we will assume the more 
likely model B to be correct. Fynbo et al. (1999a, their Table 4) 
lists the Ly$\alpha$ line luminosities of the known galaxy 
counterparts of high redshifts DLAs and LLSs, namely the DLAs towards 
PKS0528-250 (named S1), Q0151+048A (named S4) and Q2059-360, and the 
LLS towards Q2233+131. The observed Ly$\alpha$ line luminosity of S6 
is within the range 1.1--12 $\times$ 10$^{42}$ h$^{-2}$ erg s$^{-1}$ 
of these other systems.

The size of the line emission region is 6$\times$4 arcsec$^2$ 
corresponding to 22$\times$14 h$^{-2}$ kpc$^2$ at z=3.036. This size and 
the morphology of S6 are both near--identical to those of the emission 
line object S4, which has been shown in a spectroscopic study (M{\o}ller 
et al. 1998) to be a DLA galaxy unrelated to the underlying quasar.
Before the precise systemic redshifts of both the QSO and S6 have been
measured we cannot exclude that S6 is related to Q1205-30.

From neither S4, nor S6, do we detect what is obviously continuum
emission from a high redshift galaxy. This, however, is not surprising
as continuum emission from the high redshift DLA galaxy S1 was not
detected in the deep ground based images of S1 (M{\o}ller \& Warren,
1993a), but a compact galaxy was subsequently found in the HST images
(M{\o}ller \& Warren 1998). Since the impact parameters of S1, S4,
and S6 are similar, clearly HST imaging is required to resolve the
question of their continuum morphology.

Both the comparatively large luminosity and the extended morphology
of S6 is in contrast to six faint and compact emission line galaxies
(S7--S12) found at much larger distances from the quasar.
We hence consider it likely, as in the case of Q2059-360 (Leibundgut \&
Robertson 1998, Fynbo et al. 1999a), that the emission
from S6 is significantly influenced by the proximity of the active
nucleus. It is, however, possible that gravitational lensing by 
the red galaxy g2, if it is a foreground elliptical galaxy, makes S6 
appear more extended and increases
its observed flux. Here we estimate the likely strength of this effect.

Assuming that g2 is at z=0.5 and has an isothermal mass profile 
with a 
velocity dispersion of 300 km s$^{-1}$, the radius of its Einstein ring
is given by 
\[
\theta_E = 4 \pi (\frac{\sigma}{c})^2 \frac{d_{LS}}{d_S} = 1\farcs6, 
\]
where $d_{LS}$ and ${d_S}$ are the proper angular diameter distances 
between g2 (Lens) and S6 (Source) and between the earth and S6 
respectively, in the assumed cosmology. Gravitational lensing in this 
simple model moves S6 and Q1205-30 by an angle $\theta_E$ in the radial 
direction away from g2 causing a tangential stretching of S6 as well as an 
increase of the observed impact parameter of S6 relative to Q1205-30 by 
a factor of $\approx \theta_{g2} / (\theta_{g2}-\theta_E)$=2.3 (where 
$\theta_{g2}$=2.8 arcsec is the observed impact parameter of g2 relative 
to the QSO). The magnification of the flux of S6 would be in the range 
2--10 depending on the unknown size and orientation of the diamond
caustic of g2. A possible counter image on the eastern side of g2 would 
make S6 look more extended in the relatively low resolution of the narrow 
band image, in the direction of the observed elongation which would be 
of order 2$\theta_E$.

\subsection{The field population of Ly$\alpha$ emitting galaxies}

The presence of the six emission line galaxies S7--S12 in the field 
of Q1205-30 is interesting for several reasons. The faintness of the
broad band magnitudes of the objects makes it unlikely that they are low
redshift (z $\approx$ 0.31) galaxies with Oxygen emission lines in the 
narrow filter.
The most likely identification of the emission lines is Ly$\alpha$
at the same redshift as the Q1205-30 LLS absorber, and in what follows
we shall assume this to be the case.

Only one of the six emission line galaxies (S9) would have been
selected as an LBG in current ground based programmes. The remaining five 
are too faint in the broad bands (see Table 4). 
The number of LBGs brighter than R(AB)=25.5 with redshifts between 3.0 
and 3.5 selected from ground based surveys is 0.4$\pm$0.07 galaxies 
arcmin$^{-2}$ (Steidel et al. 1996), and we hence expect 11 in our 
27.6 arcmin$^{2}$ field. Assuming that LBGs have redshifts uniformly 
distributed between 3.0 and 3.5 we would expect on average 0.4 LBGs 
within the redshift range $dz=0.016$ corresponding to the width of the
narrow filter. In the Hubble Deep Field North the LF of LBGs has been 
extended to R(AB)=27 (Steidel et al. 1999). Integrating this LF leads 
to an expected number of LBGs of about 0.3 galaxies arcmin$^{-2}$ 
within the redshift range $dz=0.016$ or on average 9 in our 27.6 
arcmin$^{2}$ field (see Sect. 4.2 in Fynbo et al. (1999a)
for a discussion on the relation between the redshift density
$\frac{dN}{dz}$ of DLAs and the LF of LBGs). Hu et al., 
1998, have reported on detections of Ly$\alpha$ emitting galaxies by 
means of narrow band imaging and long slit spectroscopy in random 
fields with densities about 4 arcmin$^{-2}$ per unit redshift down 
to similar depths in observed flux as in our sample, but at 
somewhat higher redshifts.

Hence, narrow/broad band Ly$\alpha$ imaging indicate the existence 
of a significant population of high redshift galaxies that is not 
included in current ground based LBG samples and which is possibly
identical to the faint LBGs detected by HST. Those galaxies, which 
make up the faint end of the high redshift galaxy LF, have large 
Ly$\alpha$ equivalent widths and are about a factor of 10 more numerous 
than the bright LBGs of current ground based samples. This result 
is in good agreement with our earlier conclusion based on searches 
for DLA galaxies (M{\o}ller \& Warren 1998; Fynbo et al., 1999a) that 
there is a significant population of faint high redshift galaxies 
which, despite their very small HI absorption cross--section, 
make up most of the high redshift DLA absorbers. 

Total Ly$\alpha$ luminosities for S7--S12 are in the range
3.3--9.5 $\times$ 10$^{41}$ h$^{-2}$ erg s$^{-1}$ for $\Omega_m$=1.0 and
12--34 $\times$ 10$^{41}$ h$^{-2}$ erg s$^{-1}$ for $\Omega_m$=0.
Using the Kennicutt (1983) prescription SFR =
L(H$\alpha$)/1.12 $\times$ 10$^{41}$ erg s$^{-1}$ and assuming
L(Ly$\alpha$)/L(H$\alpha$)=10 and negligible dust extinction, we find
star formation rates in the range 0.2 -- 0.8 h$^{-2}$ M$_{\sun}$ \
yr$^{-1}$ for $\Omega_m$=1.0 and 1.1--3.0 h$^{-2}$ M$_{\sun}$  yr$^{-1}$ for 
$\Omega_m$=0. The large equivalent widths of the emission lines make strong 
dust obscuration unlikely. The star formation rates for R(AB)$<$25.5 LBGs 
as estimated from Balmer and OII emission line strengths by Pettini et al., 
1998, fall in the range 20 -- 270 M$_{\sun}$ yr$^{-1}$ for $\Omega_m$=0.2 and 
h=0.7. For $\Omega_m$=1.0 this corresponds to 4--55h$^{-2}$ M$_{\sun}$ 
yr$^{-1}$. Hence, the Ly$\alpha$ emitting galaxies have star formation 
rates about an order of magnitude lower than what is measured for the 
bright LBG population. Nevertheless, since the population of
Ly$\alpha$ emission line galaxies is much more common that the brighter
LBG population, it is still an open question if the {\it integrated}
SFR is dominated by the faint or the bright end of the high redshift
galaxy LF.

As seen in Fig.~\ref{field} the faint galaxies are all located in a
region north west of the QSO. Their projected impact parameters from
Q1205-30 range from 156 to 444 h$^{-1}$ kpc, and four of the six
galaxies are found in a small region of projected size
100 $\times$ 100 h$^{-2}$ kpc$^{2}$. As for the compact group S1--S2--S3
associated with a z=2.81 DLA absorber,
this compact group of faint galaxies associated with a z=3.036 LLS
absorber is destined to merge on a short timescale
(Navarro et al. 1995; Haehnelt et al. 1998). This, then, supports the
suggestion (Warren \& M\o ller 1996; M{\o}ller \& Warren 1998)
that high redshift, high column density absorbers predominantly are
proto--galactic sub--clumps in the process of early galaxy assembly,
rather than fully formed large rotating disks.

\section*{Acknowledgments}
We wish to thank the NTT service observation team for providing us
with these excellent data. We thank Jacqueline Bergeron, Tom 
Broadhurst and Stephen J. Warren for numerous discussions and several
useful comments on an earlier version of this manuscript. J.U.F. wishes 
to thank the UK Schmidt Telescope Unit for generous help in the effort 
of locating Q1205-30.

\end{document}